\documentclass[aip,reprint,amsmath,amsmath,amsfonts]{revtex4-1}

\usepackage{amssymb}
\usepackage{amsmath}
\usepackage{bm}
\usepackage{paralist}
\usepackage{graphicx}
\usepackage{enumerate}
\usepackage{tikz}
\usepackage{graphicx}
\usepackage{verbatim}
\usepackage{etoolbox}
\usepackage{nicematrix}
\usepackage{float}
\usepackage{setspace}
\usepackage{subcaption}

\usetikzlibrary{matrix,positioning,fit}

\usepackage{hyperref}
\hypersetup{
	colorlinks,
	citecolor=blue,
	filecolor=blue,
	linkcolor=blue,
	urlcolor=blue,
	pdfproducer={}
}

\let\bbordermatrix\bordermatrix
\patchcmd{\bbordermatrix}{8.75}{4.75}{}{}
\patchcmd{\bbordermatrix}{\left(}{\left[}{}{}
\patchcmd{\bbordermatrix}{\right)}{\right]}{}{}

\newlength{\Mylen}

\usepackage{ragged2e}
\usepackage{caption}
\usepackage{ragged2e}
\DeclareCaptionJustification{justified}{\justifying}
\usepackage{soul}	

\usepackage [english]{babel}		
\usepackage [autostyle, english = american]{csquotes}
\MakeOuterQuote{"}

\usepackage[normalem]{ulem}

\usepackage[percent]{overpic}

\usepackage{xcolor}

\usepackage{cprotect}

\makeatletter
\def\@email#1#2{%
 \endgroup
 \patchcmd{\titleblock@produce}
  {\frontmatter@RRAPformat}
  {\frontmatter@RRAPformat{\produce@RRAP{*#1\href{mailto:#2}{#2}}}\frontmatter@RRAPformat}
  {}{}
}%
\makeatother

\begin{document} 

\title{Computing Classical Escape Rates from Periodic Orbits in Chaotic Hydrogen}

\author{Ethan T. Custodio}
\email{ecustodio@ucmerced.edu}
\affiliation{Physics Department, University of California, Merced, CA
   95344, USA}

\author{Sulimon Sattari}
\affiliation{Research Institute for Electronic Science, Hokkaido University, Sapporo, Hokkaido 0010020, Japan}

\author{Kevin A.~Mitchell}
\email{kmitchell@ucmerced.edu}
\affiliation{Physics Department, University of California, Merced, CA
   95344, USA}

\date{\today}
	
\begin{abstract}
    When placed in parallel magnetic and electric fields, the electron trajectories of a classical hydrogen atom are chaotic. The classical escape rate of such a system can be computed with classical trajectory Monte Carlo techniques, but these computations require enormous numbers of trajectories, provide little understanding of the dynamical mechanisms involved, and must be completely rerun for any change of system parameter, no matter how small. We demonstrate an alternative technique to classical trajectory Monte Carlo computations, based on classical periodic orbit theory. In this technique, escape rates are computed from a relatively modest number (a few thousand) of periodic orbits of the system. One only needs the orbits' periods and stability eigenvalues. A major advantage of this approach is that one does not need to repeat the entire analysis from scratch as system parameters are varied; one can numerically continue the periodic orbits instead. We demonstrate the periodic orbit technique for the ionization of a hydrogen atom in applied parallel electric and magnetic fields.  Using fundamental theories of phase space geometry, we also show how to generate nontrivial symbolic dynamics for acquiring periodic orbits in physical systems.  A detailed analysis of heteroclinic tangles and how they relate to bifurcations in periodic orbits is also presented. 
\end{abstract}

\maketitle

\begin{quotation}
Periodic orbits are special trajectories in nonlinear systems that form closed cycles. That is, after some finite time the trajectory will begin to retrace itself. Unstable periodic orbits have a neighborhood around them over which they exert some dynamical influence quantified by their Lyapunov exponent. Any trajectory in the system can be thought of as moving from the neighborhood of one unstable periodic orbit to another until it becomes trapped or escapes to infinity. Thus one can use these orbits as a ``skeleton'' to compute dynamical averages without the need for statistical simulation techniques like Monte Carlo methods. Here we will use the method of periodic orbits to compute escape rates based on a classical atomic Hamiltonian.
\end{quotation}

\section{Introduction}
The importance of periodic orbits for characterizing classical chaotic dynamics was first realized by Poincar\'{e}~\cite{Poincare90}, who understood periodic orbits as a skeleton of the phase space dynamics~\cite{Cvitanovic91}. Since Poincar\'{e}, a rich and well developed theory has developed that shows how periodic orbits can be used to compute and characterize the statistical behavior of classical and quantum dynamical systems~\cite{cvitanovicChaosClassicalQuantuma}.  The appeal of such periodic orbit theory is that it reduces a complex system to a set of prototypical dynamical behaviors (the periodic orbits).  The quantum problem has perhaps received the most attention.  Gutzwiller, in his seminal work~\cite{Gutzwiller71,Gutzwiller90}, showed that fluctuations in the spectral density of a quantum system were attributable to periodic orbits of the corresponding classical system.  This ultimately led to significant new insights into the role periodic orbits play in physical applications, such as the absorption spectra of highly excited atoms (i.e. Rydberg atoms) in applied fields~\cite{Holle86,Main86,Holle88,Du88,Du88b,Main94}.

Though most applications of periodic orbit theory in quantum systems have looked at oscillations in the density of states, in a few select cases, periodic orbit theory could be pushed further to resolve energy levels (or resonances) of chaotic spectra as a sum over contributions from many periodic orbits~\cite{cvitanovicChaosClassicalQuantuma}. To date, we know of two well developed examples of this: the three-disk scatterer~\cite{Cvitanovic89} and the one-dimensional helium atom~\cite{Wintgen92}.  The absence of more examples highlights a lack of understanding of the conditions under which periodic orbit theory can be successfully applied to resolve chaotic spectra.

In addition to quantum applications, classical phase space averages can be computed from sums over periodic orbits.  Again, the three-disk scatterer is a prototypical example, in which the escape rate of trajectories trapped between the three disks can be computed from a sum over periodic orbit contributions~\cite{Cvitanovic94}. But there are still unresolved issues regarding how broadly such techniques can be applied.  For example, despite some prior work~\cite{Sattari17}, the question of how to apply periodic orbit theory to mixed phase spaces remains an open challenge.

There has recently also been a surge of interest in periodic orbit theory applied to high-dimensional phase spaces and dynamical systems defined by partial differential equations.  Most notable here is the success in understanding the transition to fluid turbulence via periodic orbit decompositions of solutions to the Navier-Stokes equation at intermediate Reynolds number\cite{Gibson08,Budanur17,Graham21,Avila23}.  The hope is to eventually compute statistical averages of turbulent motion with a small number of periodic orbits capturing the essential features of turbulence. Higher dimensional turbulent systems rarely have a finite grammar which completely describes phase space transport. Thus, computing a full set of periodic orbits in such a system is an impossible task. In these systems a truncated set of periodic orbits computed based on stability can be used to compute dynamical averages\cite{cvitanovicChaosClassicalQuantuma}.

As exciting as the recent high-dimensional developments are, there are still relatively few low-dimensional physical examples of the application of periodic orbit theory, especially those with quantitatively accurate periodic orbit computations. To help fill in this void, the current paper presents a highly accurate periodic orbit computation for the classical decay of a hydrogen atom in parallel electric and magnetic fields. We use the theory of heteroclinic tangles to create a Markov partition of phase space that completely describes phase space transport. Using this partition, we compute a complete set of periodic orbits up to a given discrete-time period. Then using the spectral determinant form of periodic orbit theory, we compute the classical decay rate across a wide range of parameter values, which span so-called hyperbolic plateaus. Utilizing the theory of heteroclinic tangles we are able to identify and compute the boundaries of two hyperbolic plateaus.  The transition from one plateau to the other requires pruning of periodic orbits as the symbolic dynamics of the system changes.  

While this paper presents a relatively simple case with only two fixed points and a modest Hamiltonian, these methods can also be applied to more complicated two-dimensional systems. Homotopic Lobe Dynamics (HLD) can be used to partition phases with any number of fixed points and nested tangles so long as there is a finite symbolic dynamics\cite{mitchellPartitioningTwodimensionalMixed2012}. As system complexity increases additional a-priori knowledge becomes necessary to locate all the fixed points and find parameter values that have finite symbolic dynamics. The time it takes to compute all periodic orbits will certainly increase with complexity, but they only need to be computed once and then can be numerically continued through parameter space.

The paper is organized as follows.  In Sect.~\ref{II} the system and its Hamiltonian are introduced. In Sect.~\ref{III} we compute the escape rate from a classical trajectory Monte Carlo simulation. Section~\ref{IV} presents a discussion of periodic orbit theory and how spectral determinants can be used to compute the escape rate. At this point the discrete dynamics are introduced. Section~\ref{V} defines a surface of section in order to define a discrete-time map. In Sect.~\ref{VI} a method for computing periodic orbits is presented by creating Markov partitions of phase space using heteroclinic tangles. Section~\ref{VII} uses the periodic orbits from the previous section to compute the escape rate with spectral determinants. Section~\ref{VIII} discusses heteroclinic tangles in more detail and how their topology changes as parameters are varied. Two hyperbolic plateaus are identified using heteroclinic tangencies to define the borders. Finally, in Sect.~\ref{IX} we numerically continue the periodic orbits to fully explore both hyperbolic plateaus.

\section{Hydrogen in Parallel Electric and Magnetic Fields} \label{II}

We consider hydrogen in parallel electric and magnetic fields because it retains an axis of symmetry, which reduces it to a two degree-of-freedom Hamiltonian system with chaos.  The classical Hamiltonian of the hydrogenic electron, with fields aligned along the $z$ axis, is
\begin{equation}
    H\left( \rho, z, p_\rho, p_z \right) = \frac{1}{2}\left( p_\rho^2 + p_z^2 \right) -\frac{1}{\sqrt{\rho^2 + z^2}} + Fz + \frac{1}{8}B^2\rho^2,
\label{r1}
\end{equation}
where $(\rho, z)$ are cylindrical coordinates and $(p_\rho, p_z)$ are their conjugate momenta. This Hamiltonian is expressed in a frame rotating about the $z$ axis with frequency $\omega = B/2$ to eliminate the term linear in $B$. Furthermore, we have assumed the angular momentum along the $z$ axis vanishes, i.e. $L_z = 0$. 
 As is standard~\cite{Gao92a,Gao92,Gao94} the variables $\left(\rho, z, p_\rho, p_z \right)$ are scaled by the electric field $\Tilde{F}$ according to $\left( \rho, z \right) = \left( \Tilde{\rho}\Tilde{F}^\frac{1}{2}, \Tilde{z}\Tilde{F}^\frac{1}{2} \right)$ and $\left( p_\rho, p_z \right) = \left( \Tilde{p}_\rho \Tilde{F}^{-\frac{1}{4}}, \Tilde{p}_z \Tilde{F}^{-\frac{1}{4}} \right)$. The magnetic field and electron energy are scaled like $B = \Tilde{B} \Tilde{F}^{-\frac{3}{4}}$ and $E = \Tilde{E} \Tilde{F}^{-\frac{1}{2}}$. Here, the tilded symbols represent the physical, unscaled variables and the regular untilded symbols represent the scaled variables. \cite{mitchellAnalysisChaosinducedPulse2004}
  
As is common in the literature \cite{haggertyRecurrenceSpectroscopyTimedependent2000}, we next transform into parabolic coordinates $(u,v)$ defined by
\begin{equation}
    u = \pm \sqrt{\rho + z} \, , \quad v = \pm \sqrt{\rho - z},
\end{equation}
with conjugate momenta
\begin{equation}
    p_u = v p_{\rho} + u p_z \, , \quad p_v = u p_{\rho} - v p_z.
\end{equation}
Finally, the Hamiltonian (\ref{r1}) is transformed into $h = 2r \left( H - E \right)$, where $r = \sqrt{\rho^2 + z^2}$, to remove the Coulomb singularity at the origin, so that
\begin{equation}
    h(u,v,p_u,p_v) = \frac{1}{2}(p^2_u + p^2_v) + V\left(u, v\right)
\label{hamiltonian}
\end{equation}
and
\begin{equation}
    V\left(u, v\right) = - E(u^2 + v^2) + \frac{1}{8}B^2(u^2v^4 + u^4v^2) + \frac{1}{2}(u^4 - v^4) - 2.
\label{potential}
\end{equation}
We recover $H = E$ by requiring $h = 0$.  The electron energy $E$ now behaves as a parameter of our new Hamiltonian $h$. The other parameter $B$ acts as a coupling constant between the spatial coordinates, which generates the chaotic mixing we are interested in studying \cite{mitchellAnalysisChaosinducedPulse2004}.  Note that the new Hamiltonian $h$ has a conjugate time variable $s$ defined by $ds/dt = 1/(2r)$.  In the following, all references to the continuous time variable will refer to the new time $s$.

\section{Escape Rate from Classical Trajectory Monte Carlo} \label{III}

\begin{figure}
    \centering
    \includegraphics[width=1\linewidth]
    {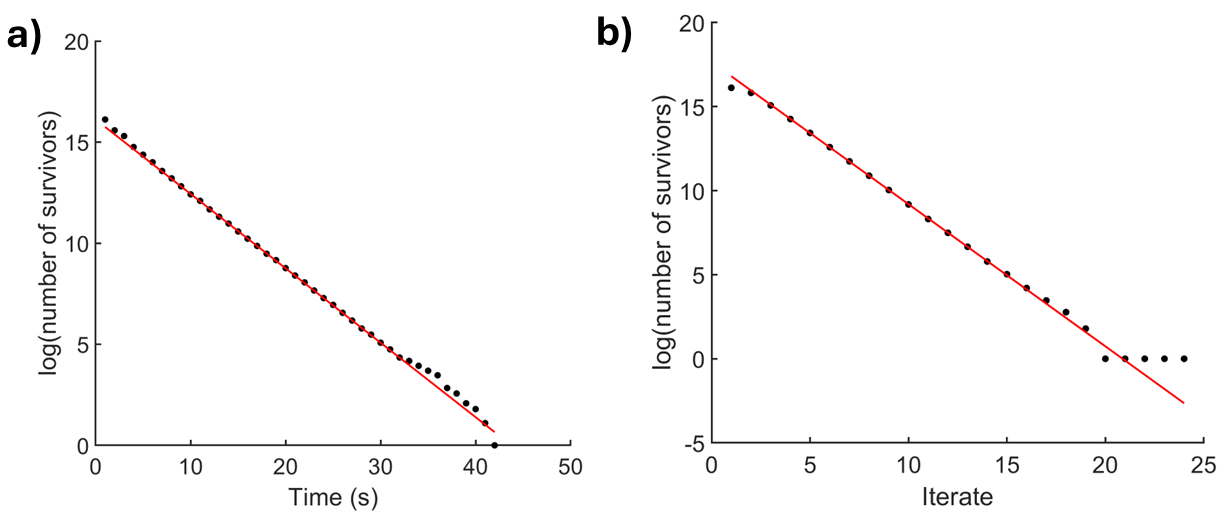}
    \caption{Exponential decay of the surviving (unionized) trajectories from classical trajectory Monte Carlo data at $E = 1$ and $B = 3.5$. Simulated data shown in black and linear fit shown in red. (a) Continuous-time simulation; $\gamma = 0.3682 \pm 0.005$. (b) Discrete-time simulation using the map; $\gamma_d = 0.8456 \pm 0.012$.}
    \label{fig:monteClassicall}
\end{figure}

We conduct classical trajectory Monte Carlo simulations to compute the classical escape rate $\gamma$ at given values of $B$ and $E$. We use an initial ensemble of $10^7$ trajectories radially propagating from the origin; this constitutes a classical model of the quantum electron state immediately after it absorbs a short laser pulse~\cite{mitchellAnalysisChaosinducedPulse2004}. Ionizing electron trajectories escape along the negative $z$ direction, and we thus define escape, i.e. ionization, to be when a trajectory reaches $z = -1$.  Once trajectories cross this escape boundary, they will continue to infinity.  
Figure~\ref{fig:monteClassicall}a plots the number of surviving (unionized) trajectories as a function of the time $s$ computed at $E = 1.0$ and $B = 3.5$.  It shows a clear exponential decay as $e^{-\gamma s}$, with decay rate $\gamma = 0.3682 \pm 0.005$ in units of inverse time.

We now discuss the fitting method for finding $\gamma$ in more detail.  At early times, many trajectories escape very quickly leading to transient behavior that overestimates the decay rate. At late times, there is simply not a statistically significant number of surviving trajectories. The proper fitting region lies somewhere between these two poorly behaved regions. To address this we use a fitting method described in Ref.~\onlinecite{deshmukhUsingScalingregionDistributions2023}. The procedure is to first generate a linear fit between every possible pair of endpoints, and then generate a histogram of the slopes from those fits. Finally, we fit a smooth probability distribution to the histogram and extract the local maximum (for the final decay rate) and standard deviation (for the error on the decay rate). This method provides two advantages: 1. It automatically computes the slope in the region of interest, without throwing away any of the dataset, while simultaneously providing a robust error measurement. 2. It automatically detects regions with multi-exponential decay, which have multiple local maxima corresponding to different decay rates. 

A problem with the Monte Carlo computation is that it does not elucidate any information about the underlying dynamics of the system. Additionally, the computation must be entirely recomputed for even a small change in parameter values. We will introduce an alternative approach using periodic orbits in the next section and use the Monte Carlo data to verify our results.

\section{Periodic Orbit Theory and Spectral Determinants} \label{IV}

Here, we discuss an alternative technique to compute decay rates based on far fewer trajectories and a deeper understanding of the underlying electron dynamics.  For details, see Ref.~\onlinecite{cvitanovicChaosClassicalQuantuma}.
 We begin by considering a general two-degree-of-freedom classical Hamiltonian system with four-dimensional phase space variable $x$.  We define the operator $\mathcal{A}$, which generates time-evolution by evolving a smooth state space density $\rho \left(x, t\right)$ forward in time, according to the evolution equation 
\begin{equation}
    \left( \frac{\partial}{\partial t} - \mathcal{A} \right) \rho\left( x, t\right) = 0.
\end{equation}
This motivates the definition of the spectral determinant $\text{det} \left(s - \mathcal{A} \right)$, which we view as a function of $s$. The zeros of this determinant give the spectrum of the time-evolution generator $\mathcal{A}$~\cite{cvitanovicChaosClassicalQuantuma}, and the smallest eigenvalue determines the long-time escape rate of the system.  Furthermore, this spectral determinant can be computed as a sum over the periodic orbits of the dynamical system as follows
\begin{equation}
    \text{det}(s - \mathcal{A}) = 
    \text{exp}\left(-\sum_p\sum^{\infty}_{r=1} \frac{1}{r}
    \frac{e^{- s r T_p}}{\vert \Lambda_p\vert^{\frac{1}{2}}}\right)
    \label{specCont}
\end{equation}
where,
\begin{equation}
    \left| \Lambda_p \right| = \left| \text{det} \left(1 - M_p^r\right) \right| =\left| \left(1 - \lambda_p^r\right) \left(1 - \lambda_p^{-r}\right)\right|.
\end{equation}
Here $p$ is an index that runs through all distinct prime periodic orbits, and $r$ is an index specifying the number of times each prime orbit is retraced. (A prime orbit is one that does not retrace itself.) The matrix $M_p$ is the $2\times 2$ linearization of the prime orbit over one period (i.e. the monodromy matrix), and $T_p$ is its continuous-time period. The largest eigenvalue $\lambda_p$ of $M_p$ is used to compute the determinant in the denominator. The above periodic orbit sum can be physically interpreted as follows. The set of all non-escaping points in phase space can be densely filled by the periodic orbits.  Any long-time escaping trajectory will shadow these orbits as it escapes, shifting from the neighborhood of one unstable periodic orbit to another.  Thus properly averaging over all periodic orbits contains the same global information as the eigenvalue spectrum itself. 

The periodic orbit sum in Eq.~(\ref{specCont}) converges absolutely in the limit of including all periodic orbits.  In practice, of course, only a finite number of orbits can be computed.  Thus, to get sufficient accuracy, we compute all periodic orbits up to a reasonably high period. Computing such a large set of orbits in the full four-dimensional phase space is very challenging, so in the next section we will introduce a two-dimensional surface of section that we will use to find periodic orbits of the associated discrete mapping. The discrete orbits can be integrated to give the full continuous orbits needed here.  However, as an intermediate step, it is useful to consider the discrete map in its own right, and to study the escape rate of this map in terms of a periodic orbit sum, as discussed next.

As in the continuous case, the zeros of a spectral determinant describe the escape rate of a discrete two-dimensional map.  Now, however, the spectral determinant is written as a function of $z$ in the form $\text{det}\left(1 - z\mathcal{L}\right)$,  where $\mathcal{L}$ is the discrete time density evolution operator acting on densities in phase space.  This spectral determinant is related to periodic orbits in a similar way as the continuous case given by
\begin{equation}
    \text{det}(1 - \textit{z}\mathcal{L}) = 
    \text{exp}\left(-\sum_p\sum^{\infty}_{r=1} \frac{1}{r}
    \frac{{z^{r n_{p}}}}{\vert \Lambda_p \vert^{\frac{1}{2}}}\right),
    \label{r2}
\end{equation}
where $n_p$ is the discrete period of an orbit. The discrete and continuous cases can be related by letting $z = e^{-s}$ and $n_p = T_p$. This change of variables represents moving from using discrete iterates $n_p$ to continuous time $T_p$. 

Computing zeros of the spectral determinant directly from Eq.~(\ref{specCont}) or Eq.~(\ref{r2}) is not trivial. Here we will provide a practical description of how to perform this computation. A detailed derivation of the following method is described in Ref~(\onlinecite{cvitanovicChaosClassicalQuantuma}), and a concise derivation of the discrete case is given in Ref~(\onlinecite{Sattari17}). First, the spectral determinant up to discrete period $N$ is written as a power series expansion with coefficients $Q_n$ given by
\begin{equation}
   \left. \text{det}\left(s - \mathcal{A} \right) \right |_{N} = 1 - \sum_{n=1}^{N} Q_n z^n.
\end{equation}
The coefficients $Q_n$ are directly related to the \emph{trace coefficients} $C_n = \text{tr}\left( \mathcal{A}^n \right)$ by
\begin{equation}
    Q_n = \frac{1}{n} \left[C_n - \sum_{i=1}^{n-1} Q_i C_{n-i} \right].    
\end{equation}
Note that $Q_1 = C_1.$  Finally, the trace coefficients $C_n$ can be iteratively computed as follows
\begin{equation}
    \sum_{n=1}^{N} C_n z^n = \sum_{p} T_{p} \sum_{r=1}^{n_{p}r \le N} t_{p}^r \delta_{n_pr, N}
\end{equation}
where,
\begin{equation}
    t_{p} = \frac{z^{n_p} e^{-s T_p}}{\left| \Lambda_p \right|} .
\end{equation}
Here $t_p$ is the "local trace" associated with each $p$ cycle that acts as the weight for each periodic orbit and $n_p$ is the discrete-time period. For continuous dynamics we set $z = 1$ and find the zeros of the resulting function of $s$. For discrete dynamics we set $s = 0$ and find the roots of the resulting polynomial in $z$. Note also one must switch from continuous-time period $T_p$ to discrete-time period $n_p$.

There are several key advantages to using periodic orbit techniques over Monte Carlo methods. First, orders of magnitude fewer trajectories are needed to compute the escape rate. Monte Carlo simulations require $10^7$ trajectories while periodic orbits methods require orders of magnitude fewer. In this paper we use only $10^4$ periodic trajectories. Second, they give a more complete understanding of the underlying dynamics of the system. Decomposing the dynamics into a trace over periodic orbits provides a more robust physical framework to view the system under. Next, other bulk physical properties besides escape rates can be computed with periodic orbits such as entropies, transport coefficients, and even quantum resonances, using a semiclassical theory. Finally, as parameters are varied, periodic orbits change continuously enabling us to numerically continue orbits through parameter space rather than running an entirely new Monte Carlo simulation for each parameter value.

\section{Surface of Section and Discrete-Time Monte Carlo} \label{V}

Within the full $u v p_u p_v$ phase space we define a two-dimensional surface, called the surface of section, via the constraints $h = 0$, $u = 0$.  This surface can be parameterized by  $\left(v, p_v\right)$. An initial point in the $vp_v$ plane is then mapped forward according to $\mathcal{M}\left(v_0, \, p_{v0}  \right) = \left(v_1, \, p_{v1}\right)$. Numerically, this map takes a point $\left(v_0, \, p_{v0}  \right)$ on the surface of section with $u=0$ and $p_u = \sqrt{4 - 2 \, V(0, v) - p_v^2}$, where we have chosen $p_u > 0$.  The trajectory is thus launched into the region $u>0$.  When the trajectory intersects the surface of section again, the map returns the coordinates $\left(v_1, \, p_{v1}\right)$ of the intersection point. Note that due to symmetry, the map does not depend on our initial choice $p_u > 0$.  

All periodic orbits of the continuous-time system in the full phase space pass through the surface of section.  Thus, it is sufficient to find the periodic orbits of the discrete map $\mathcal{M}$.  Once these discrete orbits are acquired, one can recover the continuous-time orbits by integrating the discrete points through the full phase space.

\begin{figure}
    \centering
    \includegraphics[width=1\linewidth]{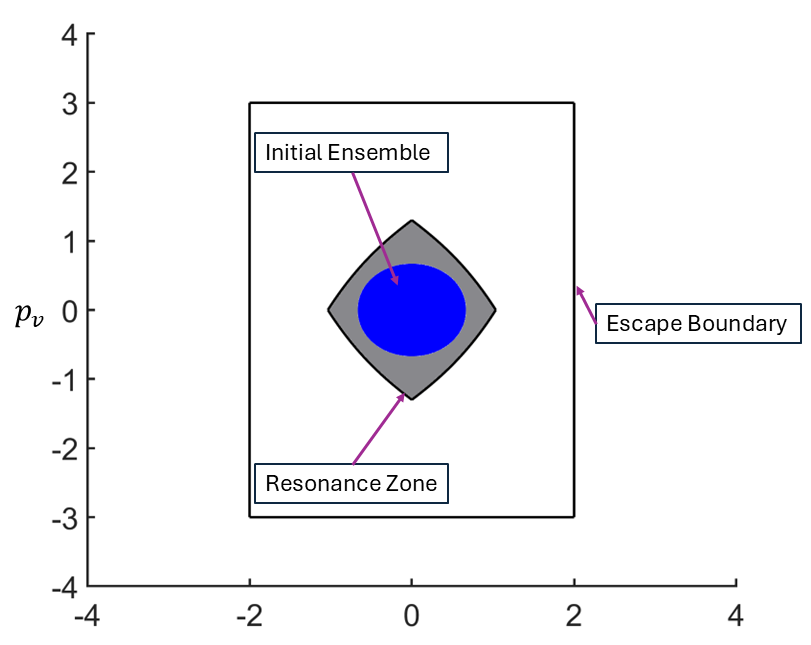}
    \caption{Initial ensemble for discrete-time Monte Carlo simulations. Initial ensemble shown in blue with $10^7$ initial points. Resonance zone shown in gray bounded by stable and unstable manifolds. (See also Fig.~\ref{fig:trellis3}a.) Escape boundary drawn as a black square around the resonance zone.}
    \label{fig:monteInit}
\end{figure}

To study the dynamics of the discrete map, we conduct discrete-time Monte Carlo simulations using $\mathcal{M}$ to compute the discrete escape rate $\gamma_d$ at given values of $B$ and $E$. We use an initial ensemble of $10^7$ points arranged in a disk in the $vp_v$ plane, centered at the origin and contained completely within the resonance zone shown in Fig.~\ref{fig:monteInit}.  When points are mapped outside the resonance zone, they will quickly escape to infinity. A box is drawn around the resonance zone, and escape is recorded when a point is mapped outside this box. Figure~\ref{fig:monteClassicall}b plots the number of surviving points as a function of iterate computed at $E = 1.0$ and $B = 3.5$. It shows clear exponential decay, with decay rate $\gamma_d = 0.8456 \pm 0.012$ in units of inverse iterate. We use the same method as the continuous-time case to generate the fit and extract the decay rate. 
This discrete decay rate can also be computed using the discrete periodic orbit method above, which we will demonstrate below in Sect.~\ref{rates}.

\section{Computing Periodic Orbits via Phase Space Partitioning} \label{VI}

\begin{figure*}
    \centering
    \includegraphics[width = 1\linewidth]
    {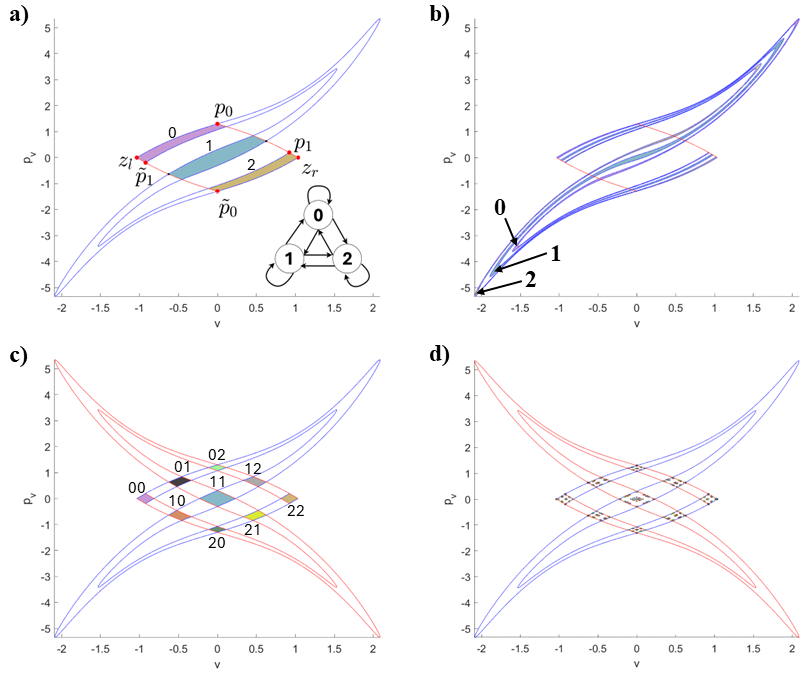}
    \caption{ Trellis at E = 1. (a) Fixed points $z_l z_r$, primary intersection points $p_0, \Tilde{p_0}, p_1, \Tilde{p_1}$, and labeled partition rectangles `0', `1', and `2'. Transition graph for transport between the rectangles shown in the bottom right. (b) The forward iterates of the partition rectangles from (a). (c) The trellis plotted with its backward iterate in red. Refined partition rectangles are colored and labeled. (d) The trellis and its backward iterate with all discrete periodic orbits up to period 6 plotted. Notice they are contained entirely within the refined partition rectangles from (c).}
    \label{fig:trellis3}
\end{figure*}

In order to compute periodic orbits, we must generate accurate initial guesses to use in a Newton's method solver. To generate this set of guesses, we will partition the $vp_v$ phase space to produce a discrete Markov process between the partition domains.  We can do this because at $E = 1$, $B = 3.5$ the dynamics on the surface of section are entirely hyperbolic with no invariant tori. There are two hyperbolic fixed points $z_l$ and $z_r$ on the $v$ axis of the surface of section, located symmetrically about the $p_v$ axis, as shown in Fig.~\ref{fig:trellis3}a. Emanating from those fixed points are one-dimensional stable (red) and unstable (blue) manifolds. All trajectories on a stable manifold are mapped towards the fixed point it emanates from. Conversely, all trajectories on an unstable manifold are mapped away from the fixed point.  The stable and unstable manifolds are infinitely long, forming a complicated structure called a heteroclinic tangle, which describes the transport of points in phase space. We will construct the phase space partition from segments of these stable and unstable manifolds.

Points where the stable and unstable manifolds intersect such that the open intervals of the manifolds up to that point do not intersect elsewhere are referred to as \textit{primary intersection points}. Fig.~\ref{fig:trellis3}a shows two primary intersection points $p_0$ and $\Tilde{p}_0$ and their forward iterates $p_1$ and $\Tilde{p}_1$. The intervals of the two stable and unstable manifolds up to the primary intersection points $p_0$ and $\Tilde{p}_0$ bound a region called the \textit{resonance zone}.  (See also Fig.~\ref{fig:monteInit}.)  Once a trajectory has escaped the resonance zone, it never returns. Physically, we interpret the resonance zone as the region where electrons are bound, or unionized. 

It is important to study the topology of the manifolds to partition phase space. Figure \ref{fig:trellis3}a shows the minimum length of the stable and unstable manifolds needed to completely describe the topology of the infinitely long manifolds. The finite length of manifolds in Fig.~\ref{fig:trellis3}a is called a trellis. A trellis is the minimum set of manifolds which contains all the topological information required to construct the partition~\cite{mitchellPartitioningTwodimensionalMixed2012}. The trellis can be used to determine a set of partition domains, or rectangles, and the symbolic dynamics between them. This can be done in a variety of ways~\cite{gonzalezDevelopmentScenarioConnecting2014}, including the method of homotopic lobe dynamics (HLD)~\cite{mitchellPartitioningTwodimensionalMixed2012}.  Here, however, we shall give an intuitive description of the partitioning and symbolic dynamics without the need for the full machinery of HLD. We first identify three topological rectangles formed by the trellis and shown shaded in Fig.~\ref{fig:trellis3}a. One can show that all trajectories that never leave the resonance zone in forward or backward time must lie within these three rectangles. The forward iterate of each of these rectangles is shown in Fig.~\ref{fig:trellis3}b, using the same coloring as in Fig.~\ref{fig:trellis3}a. Note that each iterated rectangle is stretched in length such that it passes through each of the original rectangles.  We can develop a symbolic dynamics for this process by labeling the original rectangles `0', `1', `2', as shown in Fig.~\ref{fig:trellis3}a. The iterated rectangles in Fig.~\ref{fig:trellis3}b imply that there is an allowed transition from any of the three symbols to any of the three symbols, as shown by the transition graph in Fig.~\ref{fig:trellis3}a. This transition graph defines a Markov process, known as a full shift on three symbols~\cite{JUNG1999}, that can be used to label all periodic orbits of the system. 

From the symbolic dynamics, the symbolic itinerary of every periodic orbit up to a given period can be written. Table~\ref{tab:orbits} shows the number of periodic orbits for a given length itinerary.  For a faithful symbolic representation, as we have here, we can be sure that this method captures every orbit up to the given period. For each periodic itinerary, we can find the associated periodic orbit as follows. First, the symbolic itinerary labels a sequence of partition rectangles. Next, a single representative point from each rectangle is used as an initial approximation to the periodic orbit.  Finally, this approximation is used as an initial condition in a multi-point shooting Newton's method solver to quickly converge to the periodic orbit.  For this approach to work, the initial guess must be sufficiently close to the true orbit for it to converge.  In truth, the partition rectangles shown in Figure~\ref{fig:trellis3}a are too large to provide sufficiently precise initial conditions to converge to the correct orbits. To address this issue we can refine any partition rectangle into smaller sub-rectangles with longer symbolic labels. The sub-rectangle labels correspond to both the current rectangle of a point and the next rectangle it will be mapped into. 

To refine a partition, take a partition rectangle and map it forward one iterate. Note which partition rectangles it maps into. Take a region of overlap and map it backwards into the original rectangle. The resulting rectangle is a sub-rectangle corresponding to all trajectories that map from the starting rectangle to the second rectangle. Append a symbol to this new sub-rectangle's label corresponding to the rectangle it gets mapped into. Each time this process is performed on a partition, exponentially more sub-rectangles will be generated to form a Cantor-like geometry. Figure~\ref{fig:trellis3}c shows the refined partition generated after one application of this method. Notice this has the same effect as mapping the stable manifold backward one iterate and labeling each region of overlap by the two rectangles that overlap. For any given orbit itinerary, use the center of each refined rectangle that corresponds to a substring of the orbit itinerary. For example, consider refining the `0' rectangle. It maps to itself and each other rectangle, so refining it will give us three new sub-rectangles labeled `00', `01', and `02' corresponding to each of the possible future locations of points originally within the `0' rectangle. Thus, to find the periodic orbit with the itinerary `10' choose the centers of the sub-rectangles labeled `01' and `10' to use as an initial guess. Figure~\ref{fig:trellis3}c shows the partition used to compute all periodic orbits up to $n_p = 10$. Figure~\ref{fig:trellis3}d shows all discrete orbits up to period six plotted with the trellis. 

\begin{figure*}
    \centering
    \includegraphics[width = 1\linewidth]{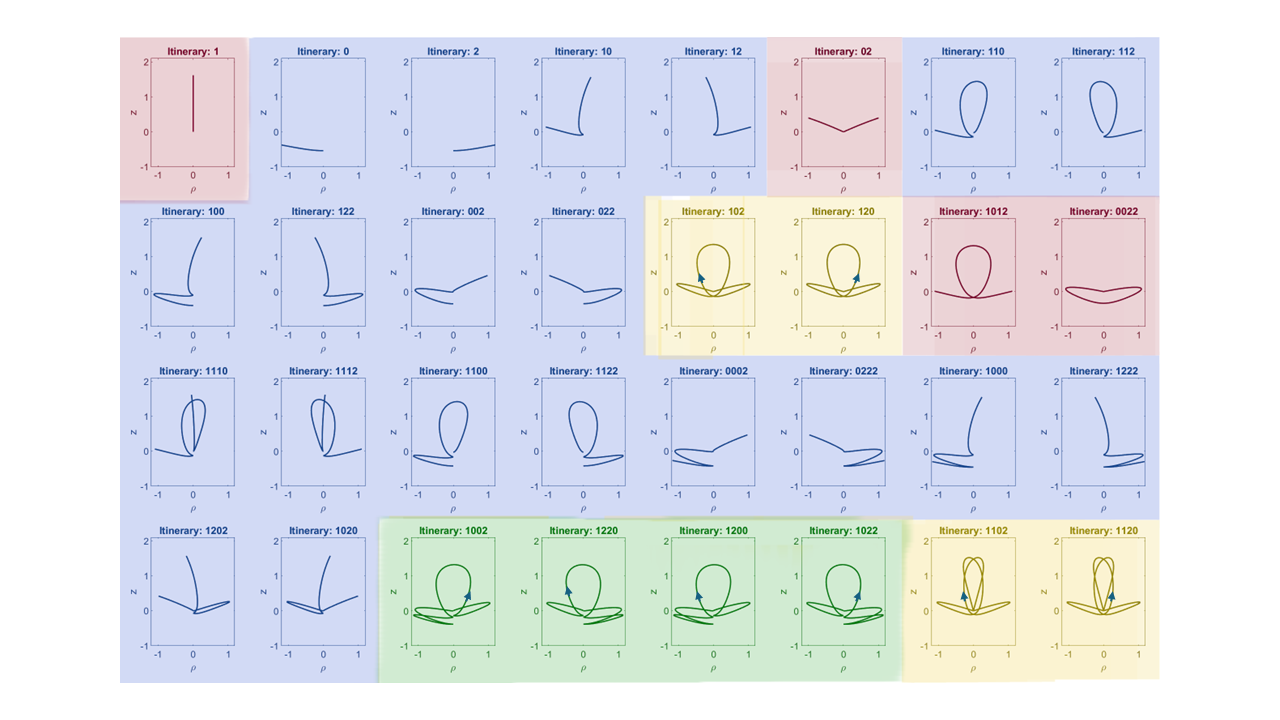}
    \caption{All continuous orbits with itinerary lengths of one through four at $E = 1$ and $B = 3.5$ plotted in the $\rho z$ configuration space. The symbolic itinerary is noted above each orbit. Notice that the itinerary length does not precisely determine the length of the continuous orbit. Orbits highlighted in blue are invariant under time reversal. Orbits highlighted in red are invariant under both time reversal and the reflection '0'$\rightarrow$'2' separately. Orbits highlighted in yellow are invariant under the composition of time reversal and reflection. Orbits highlighted in green are invariant under neither time reversal nor reflection. Arrows indicate the direction of the orbit when the trajectory does not retrace itself.}
    \label{fig:orbits}
\end{figure*}

To obtain the continuous-time trajectory in the full $u v p_u p_v$ phase space, integrate a discrete orbit forward through the full phase space. Figure~\ref{fig:orbits} shows all continuous orbits up to period four translated into the original physical $\rho z$ configuration space. We have computed a complete set of all periodic orbits, both discrete and continuous, through discrete period $n_p = 10$ for use below. We choose $n_p = 10$ because we only need to refine the partition rectangles once and there is already a reasonably large number of orbits, $\sim10^4$.

In the set of continuous periodic orbits there are two symmetries that arise from symmetries of the underlying trellis. The first is time reversal symmetry corresponding to orbits that retrace themselves. The second is reflection symmetry about the vertical axis that corresponds to exchanging the `0' and `2' symbols in an orbits' symbolic itinerary. Figure~\ref{fig:orbits} shows continuous orbits up to period four colored by their invariance under these symmetry transformations. Some orbits exhibit invariance under the pure symmetry operations, while others are invariant under both separately, the composition of the two, or neither.

\begin{table}
    \centering
    \begin{tabular}{| c | c | c |}
        \hline
        Period ($n_p$) & \# Orbits $E = 1$ & \# Orbits $E = 0.285$ \\
        \hline \hline
        1 & 3 & 3\\
        2 & 3 & 3\\
        3 & 8 & 4\\
        4 & 18 & 14\\
        5 & 48 & 44\\
        6 & 116 & 110\\
        7 & 312 & 284\\
        8 & 810 & 716\\
        9 & 2184 & 1888\\
        10 & 5880 & 4998\\
        \hline
    \end{tabular}
    \caption{Number of periodic orbits at $B=3.5$ for each discrete period up to $n_p = 10$. Values shown for $E = 1.0$ and $E = 0.285$, corresponding to the upper and lower plateau, respectively.  The values were computed directly from the transition graphs in Fig.~\ref{fig:trellis3}a and Fig.~\ref{fig:lowerTrellis}b, and they match the number of numerically computed periodic orbits exactly.}
    \label{tab:orbits}
\end{table}
\section{Escape Rate from Periodic Orbits\label{rates}} \label{VII}

\begin{figure}
    \centering
    \includegraphics[width = 1\linewidth]{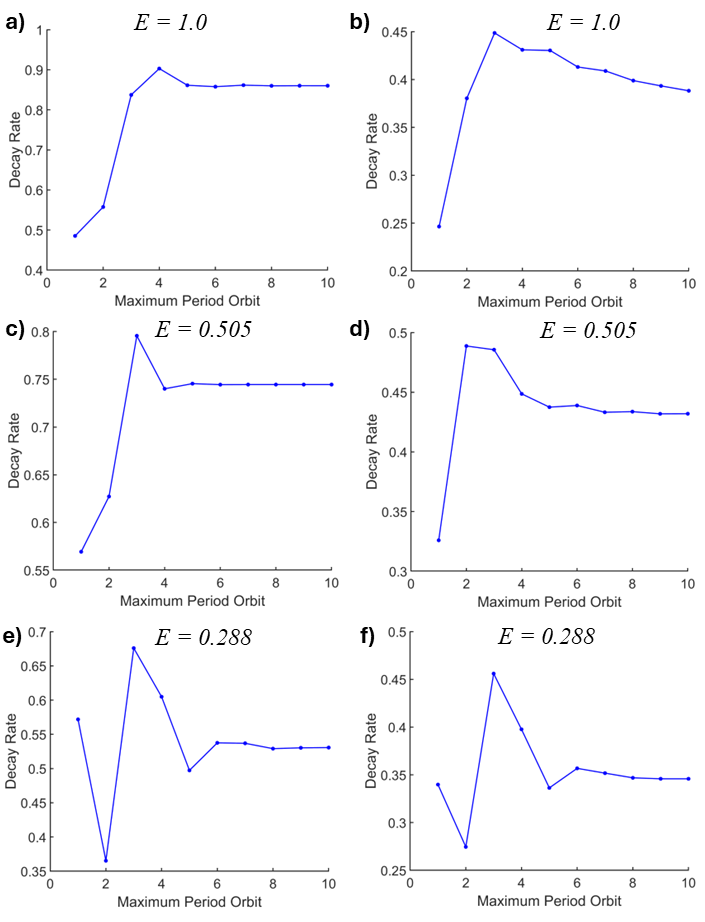}
    \caption{Decay rate computed with an increasing set of orbits. At each point all orbits up to the 'maximum period' are used to compute the decay rate. Discrete data is shown in (a), (c), and (e). Continuous data is shown in (b), (d), and (f). Each figure is computed at $B = 3.5$ with a different values of $E$ to show the difference in convergence between the regions. For the discrete case all parameter values have clearly converged. For the continuous case (b) and (d) have clearly converged while (f) has not yet converged. More orbits would be needed for convergence at that parameter value.}
    \label{fig:decayByPeriod}
\end{figure}

Now that we have computed a full set of periodic orbits up to $n_p = 10$ at $B = 3.5, E = 1.0$, we can utilize roots of the spectral determinants to compute the escape rate.  Starting with the discrete-time case, we compute the escape rate to be $\gamma_d = 0.860180 \pm 2.1 \times 10^{-5}$. This agrees well with the value computed previously from Monte Carlo data $\gamma_d = 0.8456 \pm 0.012$. Figure~\ref{fig:decayByPeriod}a shows the convergence of the discrete spectral determinant as a function of the highest period orbit used. The error is computed by taking the absolute difference between the escape rate computed with all orbits up to $n_p = 10$ and $n_p = 9$.

For the continuous case, we compute the escape rate to be $\gamma = 0.3882 \pm 5.1 \times 10^{-3}$. The error is computed in the same way as the discrete case. While the discrete case has excellent agreement with the Monte Carlo simulation, the continuous case has not quite converged. Figure \ref{fig:decayByPeriod}b shows the convergence of the spectral determinant for the continuous case. Unlike the discrete case, which has clearly converged, the continuous case requires higher period orbits to reach convergence. The convergence can be dramatically improved by decreasing the value of $E$, which we discuss in detail below.  
\section{Locating Hyperbolic Plateaus} \label{VIII}

Generally, the symbolic dynamics change as system parameters, i.e. $E$ and $B$, are varied. A region of parameter space where there exists a finite faithful symbolic representation of purely hyperbolic dynamics is called a \textit{hyperbolic plateau}. In this study we will look at escape rates computed on two hyperbolic plateaus. In this section, we discuss how the edges of these plateaus are located.

\begin{figure}
    \centering
    \includegraphics[width = 1\linewidth]
    {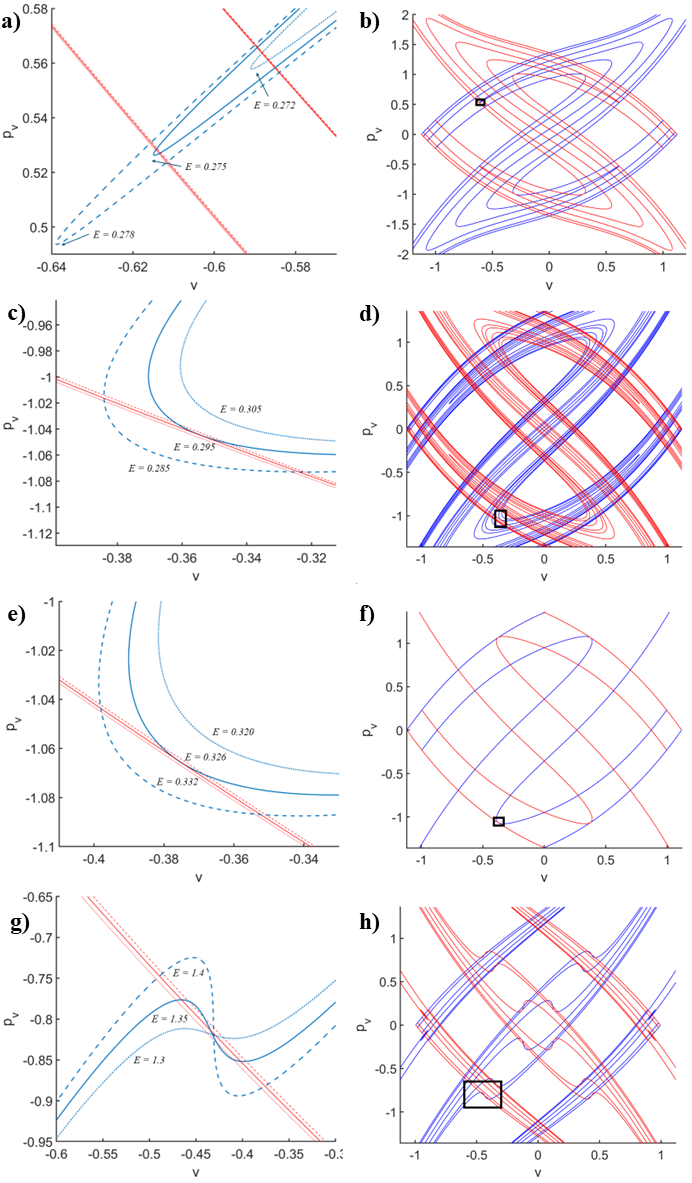}
    \caption{Tangent bifurcations as they occur near the edges of the two hyperbolic plateaus. $B = 3.5$ and $E$ is allowed to vary. Stable manifolds are plotted in red, and unstable manifolds are plotted in blue. Manifolds with the same line style are computed at the same parameter value. (a) Tangency at the bottom of the 'lower' plateau. (b) Resonance zone with zoomed in region from (a) outlined in black (c) Tangency at the top of the 'lower' plateau. (d) Resonance zone with zoomed in region from (c) outlined in black (e) Tangency at the bottom of the 'upper' plateau. (f) Resonance zone with zoomed in region from (e) outlined in black. (g) Cubic tangency at the top of the 'upper' plateau. (h) Resonance zone with zoomed in region from (g) outlined in black}
    \label{fig:tangencies}
\end{figure}

As parameters are varied, homo/heteroclinic intersection points move continuously along the stable manifolds. Two, or more, intersection points can combine in a tangent bifurcation, i.e. when the stable and unstable manifolds tangentially intersect. This breaks hyperbolicity and produces global bifurcations in the dynamics. Keeping $B$ fixed at $B = 3.5$ and varying $E$ down from $E = 1$, we observe a tangency at $E = 0.326 \pm 0.006$. This marks the lower boundary of what we will refer to as the ``upper'' plateau.  In this bifurcation, two intersection points of a ``tip'' of the unstable manifold merge and are eliminated as the tip retracts, as seen in Fig.~\ref{fig:tangencies}e. Another tangency occurs at $E = 1.35 \pm 0.05$ that marks the top boundary of the upper plateau. In this bifurcation, one intersection point splits into three as a straight segment of the unstable manifold develops a cubic oscillation.  This is shown in Fig.~\ref{fig:tangencies}g.

Ref.~\onlinecite{gonzalezDevelopmentScenarioConnecting2014} presented a general topological analysis that connects the trellis in the upper hyperbolic plateau, i.e. a trellis having a full shift on three symbols, to a different trellis defining another hyperbolic plateau. This new trellis is no longer a full shift on three symbols but is a restricted finite shift on three symbols. It thus has a smaller topological entropy. The general topological nature of this prior work implies that it should be possible to vary our parameters $E$ and/or $B$ in such a way as to identify this new trellis in the atomic system.

We hunted for this trellis by numerically computing and plotting the stable and unstable manifolds at different parameter values.   Based on the trellis topology described in  Ref.~\onlinecite{gonzalezDevelopmentScenarioConnecting2014}, we knew what tangent bifurcations would lead to the new trellis. Then, a binary search of parameter space was conducted using manifold tangencies as heuristics. Continuing to vary $E$ downward, the first tangency occurs at $E = 0.295 \pm 0.01$ marking the top border of the ``lower'' plateau, as shown in Fig.~\ref{fig:tangencies}c. Further varying $E$, the tangency that marks the lower boundary occurs at $E = 0.275 \pm 0.003$ (Figure~\ref{fig:tangencies}a).  Both of these bifurcations involve a tip retracting and two intersection points merging and disappearing.

\begin{figure*}
    \centering
    \includegraphics[width = 1\linewidth]
    {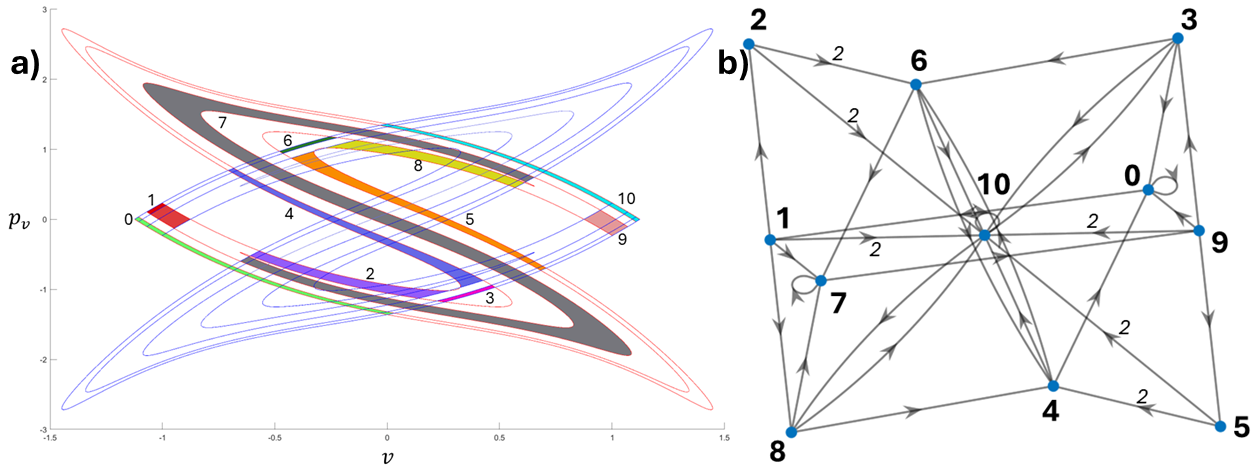}
    \caption{Symbolic dyanmics of the lower hyperbolic plateau. (a) Trellis at $E = 0.285$ with partition rectangles colored and labelled. Two forward iterates of the unstable manifold are plotted (blue) and two backward iterates of the stable manifold are plotted (red). (b) Transition graph representing the symbolic dynamics extracted from (a) using homotopic lobe dynamics.}
    \label{fig:lowerTrellis}
\end{figure*}

Figure~\ref{fig:lowerTrellis}a shows the trellis in the lower plateau, with Markov partition rectangles labelled `0' through `10'.  We identified these partition rectangles using the homotopic lobe dynamics approach in Ref.~\onlinecite{mitchellPartitioningTwodimensionalMixed2012}.  Iterating each of these rectangles forward and observing their intersections with the original rectangles, one can construct the transition graph between the Markov rectangles (as in Figs.~\ref{fig:trellis3}a and \ref{fig:trellis3}b).  Figure~\ref{fig:lowerTrellis}b shows this new transition graph. One can also use the homotopic lobe dynamics approach to directly obtain this transition graph.  

\section{Varying the electron energy} 
\label{IX}

Periodic orbits change continuously as parameters are varied, so long as they do not disappear in a bifurcation. This allows us to easily numerically continue them through parameter space. A continuation is done by making small perturbations of the parameter values and using the previously computed orbits as initial guesses for a Newton's method solver. In these cases Newton's method converges quickly because periodic orbits change continuously. 

We first continue orbits starting from $E = 1$ on the upper hyperbolic plateau up towards the top of the plateau near the cubic tangent bifurcation we observed. Next, we continue the orbits to the bottom of the upper plateau at $E = 0.326$. At this point we have periodic orbits computed across the range of the upper plateau. Next, we continue the orbits from the upper plateau down to the top of the lower hyperbolic plateau located at $E = 0.295$. A set of bifurcations occur between the plateaus that we discuss below. Finally, we continue the orbits through the lower plateau down to $E = 0.275$.

When we numerically continue orbits from the upper plateau to the lower plateau, we observe bifurcations in the set of periodic orbits. These bifurcations indicate global changes in the dynamics that correspond to topological changes in the trellis due to tangent bifurcations.  Some orbits from the upper plateau disappear before reaching the lower plateau.  However, no new orbits are created in the lower plateau.  That is, all periodic orbits in the lower plateau are a subset of the orbits in the upper plateau.

To determine which orbits to throw out and which orbits to keep, we formally continue them all to the lower plateau and then check that each of these orbits is truly a periodic orbit.  We do this by mapping an orbit forward and checking that it remains the same.  Periodic orbits that have disappeared in a bifurcation may still produce an orbit that is close to periodic, but deviates exponentially upon iteration. Orbits that deviate in this way are removed from the set of periodic orbits. From the symbolic dynamics of the lower plateau, the number of periodic orbits for a given itinerary length is known. Table~\ref{tab:orbits} shows the number of orbits on the lower plateau compared to the upper one. After the continuation and orbit pruning we find a set of orbits with the correct distribution of discrete periods predicted from the symbolic dynamics (Fig.~\ref{fig:lowerTrellis}b). 

\begin{figure}
    \centering
    \includegraphics[width=.8\linewidth]
    {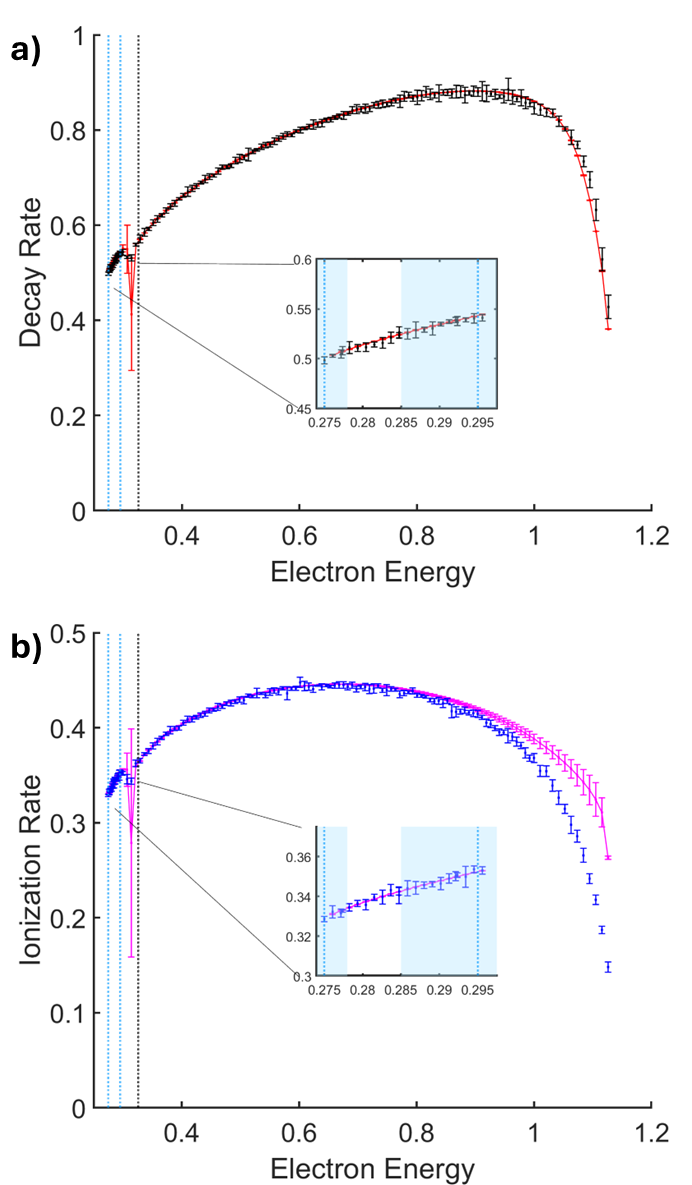}
    \caption{(a) Discrete time escape rate computed with Monte Carlo methods in black and periodic orbit methods in red. (b) Continuous time escape rate computed with Monte Carlo methods in blue and periodic orbit methods in magenta. Black vertical line marks the bottom boundary of the upper plateau. Blue vertical lines mark the boundaries of the lower plateau with error bounds illustrated by blue rectangles.}
    \label{fig:decayByE}
\end{figure}

The discrete spectral determinant computations are shown over Monte Carlo simulations in Fig.~\ref{fig:decayByE}a across both hyperbolic plateaus. The confidence intervals are computed as the difference between using all orbits up to $n_p = 10$ and $n_p = 9$. The convergence of the discrete spectral determinant at three parameter values is shown in Fig.~\ref{fig:decayByPeriod}a, Fig.~\ref{fig:decayByPeriod}c, and Fig.~\ref{fig:decayByPeriod}e. All parameter values show accurate convergence and our methods appear to be capturing the totality of the discrete dynamics. Notice the difference between the computations at the tops of each hyperbolic plateau. At the top of the upper plateau the escape rate drops precipitously towards zero. We do not see the same behavior at the top of the lower plateau. We suspect this is due to the cubic nature of the tangency on the upper plateau, but we have not precisely determined the true cause.

The continuous spectral determinant computations are shown over Monte Carlo simulations in Fig.~\ref{fig:decayByE}b across both hyperbolic plateaus. The confidence intervals are computed in the same way as for the discrete case. The convergence of the continuous spectral determinant at three parameter values is shown in Fig.~\ref{fig:decayByPeriod}b, Fig.~\ref{fig:decayByPeriod}d, and Fig.~\ref{fig:decayByPeriod}f. There is excellent convergence on the lower plateau and the bottom of the upper plateau. However, compared to the discrete case, the continuous spectral determinant at the top of the upper plateau begins to converge more slowly. This implies a larger set of periodic orbits is needed in that region to fully capture the escape dynamics. We suspect this is due to the cubic nature of the tangency at the top of the plateau. Likely long orbits that shadow heteroclinic orbits become more important as you get closer to a cubic tangency, but this is still a conjecture.

\section{Conclusion}

Periodic orbits act as the skeleton of a dynamical system, and spectral determinants can be used to compute escape rates from periodic orbits. Escape rates computed in this way may use orders of magnitude fewer trajectories than Monte Carlo simulations, in this study four orders of magnitude fewer. A second advantage to periodic orbit methods is that they do not have to be entirely recomputed for a change in parameters. In a chaotic system even a small change in parameters requires an entirely new Monte Carlo simulation to accurately compute escape. On the other hand, periodic orbits computed for one parameter can be numerically continued to another parameter value rather than recomputing all the orbits from scratch. 

Additionally, the periodic orbit methods presented in this paper allow for a thorough analysis of the system and provides insights that perhaps would not have been noticed otherwise. Periodic orbit theory provides a robust connection between symbolic dynamics and periodic orbits, which in conjunction with phase space partitions allows for the detailed probing of dynamics. Analyzing heteroclinic tangles tells us about global bifurcations, and bifurcations in periodic orbits indicate changes in phase space topology. Further analysis of symmetry and its role in bifurcations, though not mentioned in detail here, can provide yet another layer of understanding to chaotic systems. Furthermore, these methods also have applications to semi-classical analyses of open quantum systems. By including quantum properties of periodic orbits in the spectral determinant one can relate the zeros of the spectral determinant directly to quantum resonances. The breadth of systems that periodic orbit techniques can be applied to is still not known, but this paper expands the applications to additional experimentally testable systems.

\begin{acknowledgments}

The computation of the Monte Carlo simulations, Lyopanov exponents, continuous time periods, and periodic orbit continuations was done on the PINNACLES cluster: Intel-28-Core Xeon Gold 6330 2.0GHz nodes

\end{acknowledgments}

\section{Author Declarations}

\subsection{Conflict of Interest}

The authors have no conflicts to disclose

\subsection{Author Contributions}

\textbf{Ethan Custodio:} Conceptualization (equal), Data Curation (lead), Formal Analysis (lead), Methodology (equal), Resources (lead), Software (lead), Validation (equal), Visualization (lead), Writing/Original Draft Preparation (lead), Writing/Review \& Editing (equal)

\textbf{Sulimon Sattari:} Conceptualization (equal), Formal Analysis (support), Methodology (equal), Writing/Review \& Editing (equal)

\textbf{Kevin Mitchell:} Conceptualization (equal), Formal Analysis (support), Methodology (equal), Resources (support), Supervision (lead), Validation (equal), Writing/Review \& Editing (equal)

\section{Data Availability}

The data that support the findings of this study are available from the corresponding author upon reasonable request.

\bibliography{RydbergClassical2023,MyBibDeskBib}

\end{document}